# Nano Imprint Lithography on Silica Sol-gels: a simple route to sequential patterning**

*By Christophe Peroz, Vanessa Chauveau, Etienne Barthel and Elin Søndergård**

Since the pioneering work of S.Y. Chou et al.[1] Nano Imprint Lithography (NIL) has emerged as a promising technique for surface patterning, opening for numerous applications ranging from nanophotonics[2] to microfluidics[3]. NIL basically consists in the stamping of deformable surfaces or films. Preferred materials are thermoplastics[4] and UV curable resists[5]. So far, most papers report on single imprinting methods for which the same surface is imprinted only once. However, many applications such as biomimetic materials[6] and photonics[7] would benefit from a more versatile imprinting technique where successive imprinting *on the same area* with different stamps generates complex structures by the combination of simpler templates. Such *sequential patterning* strategies are an emerging topic in the field of NIL.

An interesting method was proposed by the groups of Lee[8] and Low[9]: they demonstrated that for polymethylmetacrylate (PMMA) the strain softening occurring during the first imprint step allows secondary imprints to be performed below the glass transition temperature of the polymer without relaxation of the first imprints. The method, which is based on plastic deformations, requires that high pressure is applied during the imprinting process. Moreover, when using a polymer layer, the resulting structure has a limited thermal and mechanical stability. Therefore, for improved stability, the imprinted layer is often used only as an etching mask to transfer the structures into a stable substrate. A major breakthrough for easier and faster processing, especially on brittle ceramic and glass substrates, would be a low pressure patterning technique on a directly functional thin film or resist omitting the final pattern transfer step. In that respect, inorganically cross-linked sol-gel (ICSG) resists appear as very attractive materials due to their low initial viscosity and outstanding thermal, chemical and mechanical stability[10,11]. Very few previous studies have addressed the imprinting of such materials[12,13]. In the present paper, we report the imprinting of square silica structures from simple line gratings and demonstrate how the specific thermo-rheological behavior of ICSG resists can be harnessed to form complex structures by sequential imprinting at low pressures.

Even though our goal is pure silica structures, the strategy for *thermal imprinting of ICSG resists*[14] is to start from hybrid silica precursors for a better control over the condensation[10]. A patterned and flexible stamp is pressed at ambient temperature onto the ICSG resist (Fig. 1 (A)); the system is heated at temperature $T_1$ for a period of time $t_1$ (Fig. 1 (B)), revealing the imprinted structures after stamp removal (Fig. 1 (C)). The structures formed during this single step imprinting are organic/inorganic hybrids; to turn them into pure silica, the organic moieties are oxidised by annealing at high temperatures (700° C°). This annealing is possible only if a condensation threshold has been exceeded during the imprinting step[14]; otherwise, the structures relax during annealing because the material temporarily turns into a fluid state as the temperature increases. By controlling the thermo-rheological properties of the ICSG resist we have been able to directly imprint features with high aspect ratios (>4) and limited shrinkage even after annealing[14].

The strategy for the *sequential* thermal imprinting of ICSG resists is as follows. Step 1 is identical to single step imprinting (Figure 1 (A)-(C)) and results in a set of primary structures. During Step 2 another patterned stamp is pressed onto this pre-patterned region (Fig 1 (D)) – here we use the same stamp after a in-plane rotation by 90°. The sample is then heated again at temperature $T_2$ during time $t_2$ (Fig 1 (E)), resulting in the formation of secondary structures (Fig 1 (F)). Annealing is carried out only after Step 2.

[*]   Dr C. Peroz, Ms. V. Chauveau, Dr E. Barthel, Dr E. Sondergard
      Laboratoire Surface du Verre et Interfaces
      Unité Mixte CNRS/Saint-Gobain UMR 125
      39 quai Lucien Lefranc
      F93303 Aubervilliers (France)
      E-mail: elin.sondergard@saint-gobain.com

[**]  The authors thank Dr C. Heitz, Dr J.M. Berquier, Dr V. Goletto for useful discussions on sol-gel science and L. Homo and C. Papret for technical assistance.

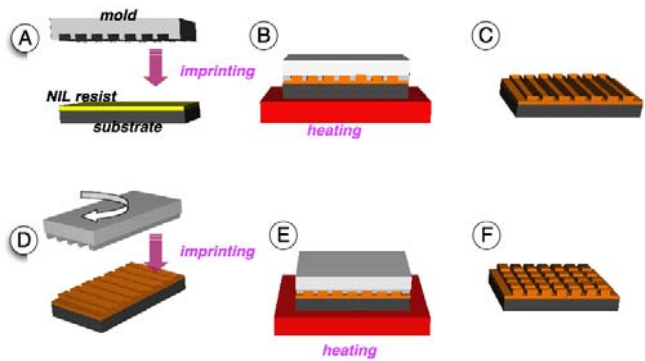

**Figure 1.** Schematic illustration of the NIL imprinting process of ICSG resists. For single step imprinting, Step 1 is performed as follows: A) the stamp is applied on the resist B) the sample is heated at temperature $T_1$ during time $t_1$ C) the stamp is separated from the sample revealing imprinted structures. For sequential imprinting Step 1 is followed by Step 2: D) another stamp is pressed onto the sample – here we use the same stamp after in-plane rotation E) the sample is heated at temperature $T_2$ during time $t_2$ F) complex structures are generated from the succession of imprints.

Effective double imprinting is achieved if the secondary structures can be formed without relaxation of the primary structures. Obviously, excessive condensation during Step 1 will prevent the formation of the secondary structures. Poor condensation will result in the relaxation of the primary structures during Step 2. Therefore successful double imprinting of ICSG resists requires accurate control of condensation during Step 1.

To that end, we performed an in depth investigation of the progress of condensation during a single imprint step. Using transmission Fourier Transform Infrared (FTIR) spectroscopy at normal incidence, we measured the time evolution of the intensity of the silanol (Si-OH) absorption peak at 895 cm$^{-1}$ for different temperatures $T_1$ in the range from 80 °C to 140 °C. The silanol groups condense into Si-O-Si siloxane bonds and the decrease of the silanol peak intensity signals the progress of the condensation[16]. In order to reproduce the thermal cycles used during the imprinting process an *in-situ* heat stage was used. The FTIR experiments were performed without the PDMS stamp. Due to high permeability of PDMS for the diffusion of small molecules including water and the initial dry state of the sol-gel film we do not expect the condensation kinetics to be altered substantially[17]. Subsequently we will use the intensity of the silanol peak (after normalization to the value at ambient temperature and denoted $\tau_{SiOH}$) to quantify the cross-linking of the ICSG resist. An example of condensation kinetics at $T_1=110$ °C is given in Figure 2. Raw spectra are displayed on Fig. 2a which depicts the marked decrease of the silanol group intensity and the appearance of distinct siloxane stretching resonances at 780 and 1020 cm$^{-1}$. Figure 2b displays the time evolution of $\tau_{SiOH}$ for different imprinting temperatures $T_1$. The region where the structures were found stable against high temperature annealing is indicated by the shaded area (Fig. 2 b). Clearly, stability is acquired at a threshold $\tau_{SiOH}$~0.3. Patterns which are imprinted under conditions where $\tau_{SiOH}<0.3$ are stable upon annealing at 700°C while they are unstable for $\tau_{SiOH}>0.3$. In the former case the condensation maintains the patterns rigid during annealing while in the latter case the film material passes into a fluid state because of insufficient cross-linking. The value of the threshold does not depend on temperature $T_1$ and appears to be a material related constant.

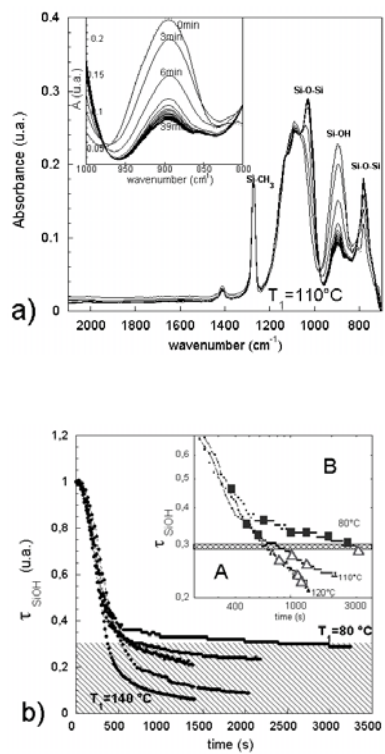

**Figure 2.** a) Time evolution of FTIR absorption spectra of ICSG resists at $T_1=$ 110 °C in the wavenumber range 2100-700 cm$^{-1}$. Spectra were taken at 3 minute intervals after heating started. Details on the time evolution of the silanol peak are reported as an inset. b) Time evolution of the intensity of the silanol peak at 895 cm$^{-1}$ normalized to its value at ambient temperature $\tau_{SiOH}$ for different temperatures $T_1=80, 110, 120, 130, 140$ °C. The shaded area indicates the region of stability of the structures against annealing. The inset brings into relation $\tau_{SiOH}$ and the possibility to perform double imprints. The symbols refer to samples where it is possible (square, zone B) to perform double imprinting or not (triangle, zone A).

This control over the thermo-rheological properties of ICSG resist has been exploited to achieve double imprinting. The vicosity acquired by the material during Step 2 must be tuned so as to: 1) prevent relaxation of the existing features and 2) allow further structuration by the PDMS stamp. As shown above, this can be achieved by adjusting the imprint parameters for Step 1 (thermal treatment temperature $T_1$ and time $t_1$). To demonstrate this possibility, double imprinting was attempted with different imprint parameters for Step 1. Specifically, square structures (Fig. 3 b) were generated from line gratings (Fig. 3 a) and we checked the feasibility of a second imprinting step as a function of $t_1$ for several temperatures $T_1$ between 80 °C and 140 °C. AFM observations were used to evaluate the quality of the final structures. The results were checked against the value of $\tau_{SiOH}$ reached at the end of Step 1, as determined by IR spectroscopy (Figure 2b, insert)). Successful double imprints are marked as filled squares while unsuccessful cases where the second imprint step did not modify the initial pattern are indicated by open triangles. The region where the second imprint is feasible coincides with the regime where a fluid state appears upon annealing at 700°C after Step 1 (Fig. 2b).

sequential imprinting the condensation of the ICSG resist during Step 1 provides a control parameter over the height of the secondary structures.

In summary, we have demonstrated that soft thermal nanoimprint of ICSG films is a powerful method to build up complex structures from simpler templates by sequential imprinting. On MTEOS resists, a material-related condensation threshold was identified. Below the threshold, a fluid state of the material is obtained upon heating which allows low pressure imprinting. By adjusting the condensation during a first imprint step, it is possible to perform double imprinting and even to tune the height of the features resulting from the second imprint step. This is illustrated by the variety of patterns which can be obtained from simple gratings, ranging from amplitude-modulated lines to dot patterns. In addition the structured surfaces benefit from the excellent stability of silica after annealing at high temperature. Enhanced knowledge about ICSG materials allows for flexible patterning techniques where omplex shapes can be obtained by superimposing simple structures. A consequence could be a generic method for multiscale structuring which would impact area as different as biomimetic materials, optical broad band or multiple wavelength applications.

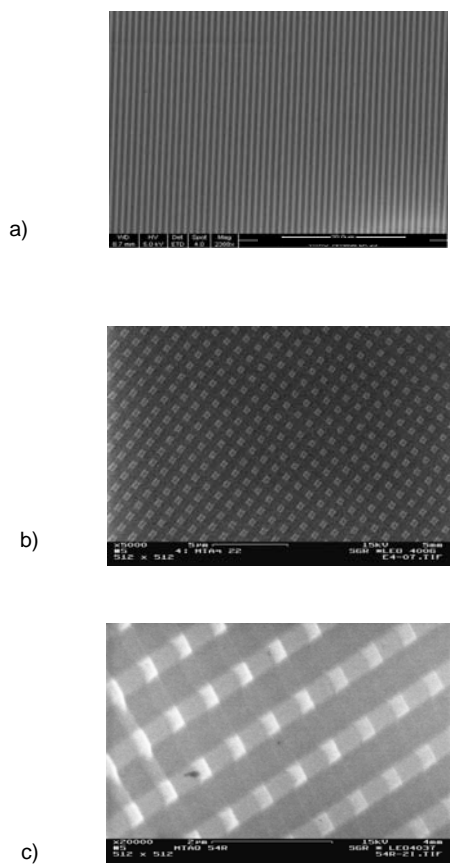

**Figure 3:** SEM pictures of gratings and square patterns imprinted in ICSG resists. (a) single step imprint of gratings for T1= 110 °C and t1=240 s; (b) a square island array resulting from an additional imprint step (T2= 110 °C, t2=240 s) after a 90° in-plane stamp rotation between the two imprints. (c) same structure after annealing to obtain a pure silica structure.

Using elaborate stamps and positioning devices would allow more complex structures to be generated. However, we have also found that the control over the thermo-rheological properties of the ICSG resists provides for fine tuning of the amplitude of the secondary structures. Again the results can be understood by considering the value of τSiOH reached at the end of bStep 1. Figure 4 depicts four samples where τSiOH after Step 1 was 0.30, 0.32, 0.33 and 0.37 respectively and the parameters for Step 2 were T2=130 °C and t2=300 s. Figure 4a shows barely noticeable relieves added to the primary imprint: close to the condensation threshold, the gel is too viscous to be affected by contact with the PDMS stamp during Step 2 and this is why the full depth of the stamp structure was not replicated. Figures 4b and c depict cases where the secondary structures have been partially imprinted to produce a ripple superimposed on the primary structures. Figure 4d displays a case where the secondary structures were completely imprinted. In this last case, the heights of the structures are identical on the template and the replica which proves that negligible relaxation of the primary structures has occurred during Step 2. The low condensation ratio of this sample maintained a state which was fluid enough to allow for a complete replication on the second template. These results clearly demonstrate that for

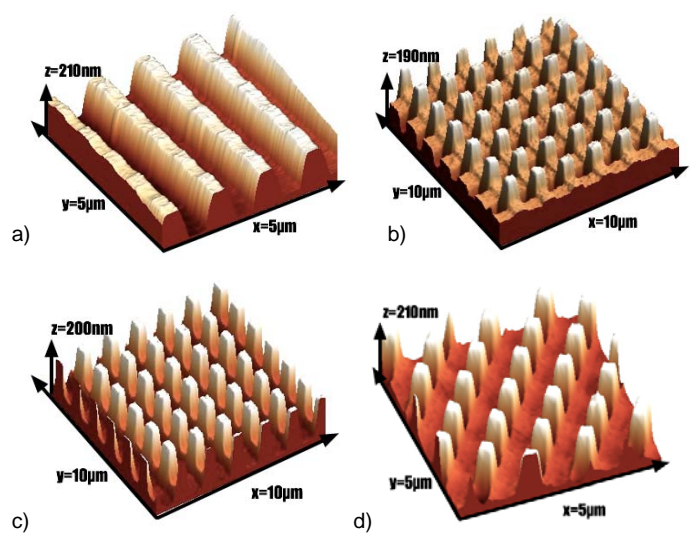

**Figure 4:** AFM images of nanostructures obtained by double imprint, with different imprint time $t_1$ at a constant temperature of $T_1$= 110 °C : the relative condensation $τ_{SiOH}$ at the end of the first imprint is a) 0.30:, b) 0.32, c) 0.33 and d) 0.37. The stamp is a line grating. It has been rotated by 90° between Step 1 and Step 2. The imprint parameters for Step 2 were T2=130°C and t2=300s. The significant difference in imprint depth is due to the change in rheology induced by the different condensation obtained during the first imprint step.

## Experimental

*Flexible stamps:* are obtained by casting liquid Polydimethylsiloxane (PDMS) on nickel or silicon nanopatterned master molds. The template consists in gratings with rectangular cross-section (340 nm linewidth, 1 µm pitch and a depth of about 160 nm). The PDMS stamps are treated by Trichloromethylsiloxane (TMCS) against adhesion.

*ICSG resist films:* The ICSG resist is prepared from a Methyltriethoxysilane (MTEOS) sol: this hybrid precursor exhibits inorganic crosslinking through siloxane bonds while the methyl group, which is non reactive, can be oxidized by thermal treatment above 550 °C. The ICSG resist is mixed in aqueous solution under acidic conditions. We use a 1:14 MTEOS/$H_2O$ molar ratio and adjust the pH at 3.1 with acetic acid for total hydrolysis of the alkoxy-silane groups and moderate condensation[15,16]. The resist is aged for 24 hours at ambient temperature before use. MTEOS films of about 300 nm thickness were deposited by spin-coating (at 3000 rpm) on silicon or glass substrates.

*Imprint process:* All imprints are carried out at low pressures (<0.2 MPa). The flexibility of the PDMS stamp allows for a good conformation to the surface.

[1] S.Y. Chou, P.R. Krauss, P.J. Renstrom,. Science 1996, 272, 85
[2] W Wu.; Z Yu, S.Y. Wang, R.S. Williams, Y.M. Liu, C Sun, X. Zhang, E. Kim, Y.R. Shen, N.X. Fang, *Appl. Phys. Lett.* 2007, 90, 063107
[3] C. Peroz, J.C. Galas, J. Shi, L. Le Gratiet, Y. Chen, *Appl. Phys. Lett.* 2006, 89, 243109
[4] M.D. Austin, H. Ge, W. Wu, M. Li, Z. Yu, D. Wasserman, S.A. Lyon, S.Y. Chou, *Appl. Phys. Let.* 2004, 84, 5299
[5] J. Haisma, M. Verheijen, K.v.D. Heuvel, J.v.D. Berg, *J. Vac. Sci. Technol. B* 1996, 14, 4124
[6] X. Gao, L. Jiang, *Nature* 2004, 432, 36
[7] R.A. Potyrailo, H. Ghiradella, A. Vertiatchikh, K. Dovidenko, J.R. Cournoyer, E. Olson, *Nature Photonics* 2007, 1, 123
[8] D.Y. Khang, H. Yoon, H.H. Lee, *Adv. Materials* 2001, 13, 749
[9] F.X. Zhang, H.Y. Low, *Nanotechnology* 2006, 17, 1884
[10] C.J. Brinker, G.W. Scherer, *Sol-Gel Science* Academic Press, San Diego CA 1990 839
[11] E. Yariv, R. Reiseld, *Optical Materials* 1999, 13, 49
[12] C. Marzolin, S.P. Smith, M. Prentiss, G.M. Whitesides, *Adv. Materials* 1998, 10, 571
[13] H. Tan, L. Chen, J. Wang, S.Y. Chou, *J. Vac. Sci. Technol. B* 2003, 21, 660
[14] C. Peroz, C. Heitz, V. Goletto, E. Barthel, E. Sondergard, *J. Vac. Sci. Technol. B* 2007, 25, L27
[15] B. Orel, R. Jese, U.L. Stangar, J. Grdadolnik, M. Puchberger, *J. Non-Cryst. Solids* 2005, 351, 530
[16] F. Brunet, *J. Non-Cryst. Solids* 1998, 231, 58
[17] E. Verneuil, A. Buguin and P. Silberzan, *Europhys. Lett* 2004, 412, 68